\documentclass[11pt,a4paper]{article}
\usepackage{amstex,amsthm,amssymb}
\usepackage{cite,indentfirst}

\def\At{{\widetilde A}}
\def\Bt{{\widetilde B}}
\def\Ct{{\widetilde C}}
\def\Dt{{\widetilde D}}
\def\Xt{{\widetilde X}}

\def\cQt{{\widetilde{\mathcal Q}}}
\def\cQ{{\mathcal Q}}

\def\Ft{{\widetilde F}}

\def\cY{{\mathcal Y}}
\def\cU{{\mathcal U}}
\def\cH{{\mathcal H}}
\def\cR{{\mathcal R}}
\def\cA{{\mathcal A}}
\def\cF{{\mathcal F}}

\def\Y{\mathbb{Y}}
\def\C{\mathbb{C}}
\def\N{\mathbb{N}}
\def\S{\mathbf{S}}

\def\perm{\mathfrak{S}}

\def\Fun{\operatorname{Fun}}
\def\Diag{\operatorname{Diag}}
\def\End{\operatorname{End}}

\def\Id{\operatorname{Id}}

\def\de{\delta}
\def\al{\alpha}
\def\La{\Lambda}

\def\hx{\hat{x}}
\def\hX{\hat{X}}

\def\sz{S^z}
\def\sp{S^+}
\def\sm{S^-}
\def\sg{\sigma}

\newtheorem{thm}{Theorem}
\newtheorem{prop}{Proposition}

\newtheorem{defi}{Definition}

\theoremstyle{remark}   
\newtheorem{rem}{Remark}
\theoremstyle{plain}

\def\Proof{\medskip\noindent {\it Proof --- \ }}

\let\qed=\cqfd

\newcommand{\ket}[1]{|\,#1\,\rangle}

\begin{document}

\begin{titlepage}
\begin{flushright}
LPENSL-TH-01/99\\
\end{flushright}
\par \vskip .1in \noindent

\begin{center}
{\LARGE Drinfel'd Twists and Functional Bethe Ansatz}\\
\end{center}
  \par \vskip .3in \noindent

\begin{center}

      {\bf V. TERRAS}
  \par \vskip .1in \noindent

{\sl  Laboratoire de Physique $^{*}$\\
Groupe de Physique Th\'eorique\\
       ENS Lyon, 46 all\'ee d'Italie 69364 Lyon CEDEX 07
       France}\\[0.6in]
\end{center}

\par \vskip .10in
\begin{center}
{\bf Abstract}\\
\end{center}

\begin{quote}
Using Functional Bethe Ansatz technique, factorizing Drinfel'd twists for
any finite dimensional irreducible representations of the Yangian
$\mathcal{Y} (sl_2)$  are constructed.
\end{quote}
\par \vskip 0.5in

\begin{flushleft}
\rule{5.1 in}{.007 in}\\
$^{*}${\small UMR 5672 du CNRS, associ\'ee \`a  l'Ecole
Normale Sup\'erieure de Lyon.}\\
{\small This work is supported by MENRT (France) fellowship AC 97-2-00119,
CNRS (France), and the EC-TMR contract FMRX-CT96-0012}.\\
{\small email: vterras\symbol{'100}enslapp.ens-lyon.fr}\\[0.2 in]

February 1999
\end{flushleft}

\end{titlepage}

\section{Introduction}

Quasi-triangular Hopf algebras provide 
a natural language for the study of low-dimen\-sional integrable systems
solvable by
the quantum inverse scattering method. 
Especially, their representations produce particular $R$-matrices solution 
of the Yang-Baxter equation~\cite{Dri85,Dri87,Jim85,Jim86}.

In~\cite{Dri83a,Dri87,Dri90a}, 
Drinfel'd introduced the notion of {\em twisting} 
for quasi-triangular quasi-Hopf algebras.
Among these Drinfel'd twists, a particular interesting class is given
by those connecting non-cocommutative coproducts to cocommutative one's.
Representation theory of 
particular examples of such
Drinfel'd twists has been studied in~\cite{MaiS96}, in terms of what is 
called there
{\em factorizing F-matrices} in the case of unitary $R$-matrices
associated to finite dimensional irreducible modules of quantum universal
enveloping Hopf algebras.
There, the corresponding $R$-matrices are completely factorized by these
$F$-matrices.
These objects have been constructed for irreducible finite 
tensor products of the fundamental
evaluation representations (spin $1/2$) 
of the Yangian $\cY(sl_2)$ and the quantum
affine algebra $\cU_q(s\widehat{l}_2)$.
They have turned out to be useful for the explicit
computation of form factors and the resolution of the quantum inverse 
problem for local spin operators (see~\cite{KitMT98})
in the case of the XXX and XXZ spin-$\frac12$ Heisenberg chains.
For the moment, no universal formula is known for such factorizing 
Drinfel'd twists for the Yangian. 

The aim of this letter is to pursue the study of representation
theory of these Drinfel'd twists, through the computation of 
factorizing $F$-matrices associated to 
any irreducible finite dimensional 
representations of $\cY(sl_2)$.
This is a generalization, to higher spin representations, 
of what has been done in~\cite{MaiS96}
for spin $1/2$, 
towards a possible universal formula.
Besides the interest in elucidating the structure of factorizing 
Drinfel'd twists,
another motivation of the present work is 
the computation of form factors, which has been achieved 
in the case of XXX or XXZ spin-$\frac12$ Heisenberg chain thanks to the 
knowledge of these $F$-matrices, and that remains 
to be done for higher spin chains. In particular, one can wonder what happens 
in the limit of infinite spin 
and for the corresponding quantum field theories.



\bigskip

Our approach is based on the Functional Bethe Ansatz (FBA) technique
developed by Sklyanin in~\cite{Skl90,Skl92} for the XXX chain.
Indeed, one can remark that the basis induced by FBA applied to operator $D$,
that is such that it is factorized as $D(u)=\prod (u-\hx_i)$ in terms
of its operator roots $\hx_i$, coincides with the $F$-basis 
of~\cite{MaiS96} for the XXX spin-$\frac12$ Heisenberg chain.
We will show that, using this FBA technique for higher spins,
we are able to construct the corresponding factorizing
$F$-matrices in this more general case by solving
only {\em linear equations}.

\section{Factorizing $F$-matrices and XXX Heisenberg chain}

For $r_i \in \N$, we denote by $V_{r_i}$ an irreducible finite dimensional
representation of the Lie algebra $sl_2$ (dim $V_{r_i}=r_i+1$), and by
$V_{r_i}(z_i)$ the corresponding evaluation representation of $\cY (sl_2)$.
It is known that every finite-dimensional irreducible $\cY (sl_2)$-module
is isomorphic to a tensor product of such evaluation 
representations~\cite{ChaP90}.

The purpose of this letter is to compute factorizing $F$-matrices associated 
to any irreducible finite dimensional
representations of the Yangian $\cY(sl_2)$, and which factorize the 
corresponding (unitary) $R$-matrices in these representations.
The notion of factorizing $F$-matrices, inspired by the representation 
theory of a particular class of Drinfel'd twists, was defined in~\cite{MaiS96}.
We recall here the basic definitions, and refer
the reader to~\cite{MaiS96} for more details.

\bigskip

The notion of factorizing twist is essentially defined for triangular
Hopf algebras $\cA$. It corresponds to invertible elements 
$\cF=\sum_i f^i\otimes f_i \in \cA\otimes\cA$ which factorize in particular
the corresponding universal $R$-matrix:
\begin{equation}
    \cR=\cF_{21}^{-1} \cF,
\end{equation}
where $\cF_{21}=\sum_i f_i\otimes f^i$.
Although Yangians are pseudo-triangular rather than triangular, it is possible
to define, for the unitary $R$-matrices associated to their finite dimensional
irreducible representations, the notion of factorizing $F$-matrix 
by analogy with
Drinfel'd twists for triangular Hopf algebras.
Such $F$-matrices are thus defined directly at the representation level
as a transcription of what happens at the universal level for triangular
Hopf algebras. To deal with such objects, we need to introduce some 
convenient notations that follow.

\begin{defi}\label{def1}
    Let $n$ be an integer and $\sigma$ an arbitrary element
  of the permutation group $\mathfrak{S}_n$.
  Let $X \in \End (V_q)$, $V_q=V_{r_1}(z_1)\otimes\dots\otimes V_{r_n}(z_n)$,
  be denoted by
  \begin{equation}\label{X}
    X_{r_1\dots r_n}(z_1,\dots,z_n)=
       \sum_i x^{(1)\, i}_{r_1}\otimes\dots\otimes x^{(n)\, i}_{r_n},
  \end{equation}
  where $x^{(j)\ i}_{r_j}\in \End(V_{r_j}(z_j))$.
  We define the extended action of the symmetry group $\mathfrak{S}_n$ on 
  $\End(V_q)$ as
  \begin{equation}\label{sg(X)}
    \sigma(X_{r_1\dots r_n}(z_1,\dots,z_n))=
    \sum_i x^{(\sigma^{-1}(1))\, i}_{r_{\sg^{-1}(1)}}\otimes\dots
        \otimes x^{(\sigma^{-1}(n))\, i}_{r_{\sg^{-1}(n)}}
      \in \End(V_{\sigma(q)}), 
  \end{equation}
  where $V_{\sigma(q)}=
  V_{r_{\sigma^{-1}(1)}}(z_{\sg^{-1}(1)})\otimes\dots\otimes 
  V_{r_{\sigma^{-1}(n)}}(z_{\sg^{-1}(n)})$.
\end{defi}

\begin{defi}\label{def2}
  Let $n\in \N$, and consider a family $\mathcal{G}_X$ of operators
  $X_{r_1\dots r_n}(z_1,\dots,z_n)$ defined for all $r_i\in \N$, 
  acting on irreducible
  tensor products of $n$ finite dimensional evaluation modules
  $V_{r_1}(z_1)\otimes\dots\otimes V_{r_n}(z_n)$. To each element $\sg$ of the
  symmetry group $\perm_n$, we associate the family $\mathcal{G}^\sg_X$ of
  operators $(X_\sg)_{r_1\dots r_n}(z_1,\dots,z_n) \in
  \End(V_{r_1}(z_1)\otimes\dots\otimes V_{r_n}(z_n))$ defined as
  \begin{equation}\label{Xsg}
     (X_\sg)_{r_1\dots r_n}(z_1,\dots,z_n)=
      \sg(X_{r_{\sg(1)}\dots r_{\sg(n)}}(z_{\sg(1)},\dots,z_{\sg(n)})).
  \end{equation}
\end{defi}

  Note that if the family  $\mathcal{G}_X$ corresponds to the different
  representations ($\rho_{r_1}\otimes\dots\otimes\rho_{r_n}$) of a
  universal operator $\mathcal{X}_{1\dots n}=
  \sum_i  x^{(1)\, i}\otimes\dots\otimes x^{(n)\, i}\in \cY (sl_2)^{\otimes n}$
  on $V_{r_1}(z_1)\otimes\dots\otimes V_{r_n}(z_n)$, the above definition
  simply means that
  \begin{align}
    (X_\sg)_{r_1\dots r_n}(z_1,\dots,z_n) &=
       \sg ((\rho_{r_{\sg(1)}}\otimes\dots\otimes\rho_{r_{\sg(n)}})
                      (\mathcal{X}_{1\dots n})),\\
        &= (\rho_{r_1}\otimes\dots\otimes\rho_{r_n})
                      (\mathcal{X}_{\sg(1)\dots \sg(n)}),
  \end{align}
  where $\mathcal{X}_{\sg(1)\dots \sg(n)}=\sg(\mathcal{X}_{1\dots n})$ 
  is the operator
  $\sum_i  x^{(\sg^{-1}(1))\, i}\otimes\dots\otimes x^{(\sg^{-1}(n))\, i}
  \in\cY (sl_2)^{\otimes n}$.

\bigskip

These notations enable us to introduce the notions of generalized $R$-matrices
associated to a given tensor product of modules and to a permutation $\sg$,
and of their factorizing $F$-matrices.

\begin{prop}\label{prop:R}
  For any integer $n$, we can define a map from the permutation group
  $\perm_n$ to $\End(V_q)$,
  $V_q = V_{r_1} (z_1 ) \otimes V_{r_2} (z_2 ) \otimes \dots 
  \otimes V_{r_n} (z_n )$,
  which associates in a unique way an element $R^\sg_q \in \End(V_q)$ to
  any permutation $\sg \in \perm_n$. It is defined recursively 
  by its values for
  simple transpositions $(i,i+1)$,
  \begin{equation}
    R^{(i,i+1)}_q = R_{r_i r_{i+1}}(z_i,z_{i+1}),
  \end{equation}
  and by the following composition law for the product of two elements of
  $\perm_n$:
  \begin{equation}
    R^{\sg_1 \sg_2}_q = (R^{\sg_2}_{\sg_1})_q \, R^{\sg_1}_q, 
                              \qquad \forall \sg_1, \sg_2 \in \perm_n.
  \end{equation}
  Here  $R_{r_i r_{i+1}} (z_i, z_{i+1})$ is  the $R$-matrix acting in 
  $V_{r_i}(z_i )\otimes V_{r_{i+1}}(z_{i+1} )$ as $R$ and as the identity 
  in all other modules in the tensor product $V_q$, 
  and $(R^{\sg_2}_{\sg_1})_{r_1\dots r_n}$ is defined from
  $R^{\sg_2}_{r_{\sg_1(1)}\dots r_{\sg_1(n)}}$ as in definition~\ref{def2}.
\end{prop}

\begin{rem}
   The consistency of this definition follows from the Yang-Baxter and
  unitary relations for the elementary $R$-matrices. It corresponds
  to representations of the intertwining relations in $\cY(sl_2)$.
\end{rem}

\begin{defi}
   By factorizing $F$-matrices is meant a family of invertible operators 
   $F$ acting on irreducible finite dimensional modules 
   $V_q = V_{r_1} (z_1 ) \otimes V_{r_2} (z_2 ) \otimes 
   \dots \otimes V_{r_n} (z_n )$, 
   defined for any integer $n$ and any $r_i\in\N$, 
   and such that for any element 
   $\sigma \in \perm_n$,
   \begin{equation}
      (F_\sg)_{r_1\dots r_n}(z_1 , \dots, z_n)\, 
      R_{r_1\dots r_n}^{\sigma} (z_1 , \dots, z_n ) = 
      F_{r_1\dots r_n} (z_1 , \dots, z_n ) ,
   \label{eq:frz}
   \end{equation}
   or in compact notations, 
   $(F_\sg)_q\, R_q^{\sigma} = F_q$, with $q = (r_1 r_2 \dots r_n)$ 
   ($F_\sg$ being given from $F$ by definition~\ref{def2}).
\label{def:factfm}
\end{defi}

In the following, for given $r_i\in\N$, we will use some simplified notations, 
denoting merely by
$X_{12\dots n}$ (instead of $X_{r_1 r_2 \dots r_n}(z_1, z_2,\dots, z_n)$)
an operator $X$ acting on the tensor product of evaluation representations
$V_{r_1} (z_1 ) \otimes V_{r_2} (z_2 ) \otimes \dots \otimes V_{r_n} (z_n )$,
and by $X_{\sg(1)\dots\sg(n)}$ the operator $X_\sg$~\eqref{Xsg} 
acting on the {\em same} tensor product (the order of spaces in the tensor
product being given {\em a priori}). 

Using these notations, we recall at last some characterizing properties of 
these factorizing
$F$-matrices which will lead to their explicit 
computation in section~\ref{sec:calculF}.

\begin{prop}
   \cite{MaiS96}
   Let $F_{1 \dots n}$ be factorizing $F$-matrices and 
   define partial $F$-matrices 
   $\Ft_{1, 2\dots n} = F_{1 \dots n}\, F_{2 \dots n}^{-1}$ 
   and $\Ft_{1 \dots n-1, n} = F_{1 \dots n}\, F_{1 \dots n-1}^{-1}$.
   They satisfy,
   \begin{align}
      \Ft_{1, 2 \dots n}\, \Ft_{2 \dots n-1, n} &= \Ft_{1 \dots n-1, n}\, 
                                    \Ft_{1, 2 \dots n-1} ,\label{cocycle} \\
      \Ft_{0, 1 \dots n} &= \Ft_{0, \sigma (1) \dots \sigma (n)}\quad 
                                        \forall \sigma \in \perm_n  , \\
      \Ft_{1 \dots n, 0} &= \Ft_{\sigma (1) \dots \sigma (n), 0} \quad 
                                         \forall \sigma \in \perm_n ,  \\
      \Ft_{1 \dots n, 0}\, F_{1 \dots n}\, R_{0 n}\dots R_{0 1} &= 
                                    \Ft_{0, 1 \dots n}\, F_{1 \dots n} .
   \end{align}
   Conversely, suppose we have defined sets of matrices $\Ft_{1, 2 \dots n}$ 
   and $\Ft_{1 \dots n-1, n}$ for any integer $n$ satisfying the above 
   properties with 
   $F_{1 \dots n} = \Ft_{1 \dots n-1, n} \dots \Ft_{12, 3}\, F_{12}$ 
   or equivalently due to the cocycle relation~\eqref{cocycle}, 
   $F_{1 \dots n} = \Ft_{1, 2 \dots n} \dots F_{n-1 n}$, 
   then these sets of matrices define factorizing $F$-matrices.
\label{prop:fundtf}
\end{prop}

\bigskip

In~\cite{MaiS96}, such factorizing $F$-matrices 
have been computed for tensor products of spin-$\frac 12$ evaluation 
representations of $\cY (sl_2)$, and it has been 
shown that, when applied to the space of states of the XXX Heisenberg 
spin-$\frac 12$ chain, they induced a new basis in which the operator entries
of the monodromy matrix have very simple forms. Here, we will reverse the 
process, computing factorizing $F$-matrices as matrices inducing a particular
change of basis in the space of states of the associated XXX chain.

Thus, for a given finite tensor product of evaluation spin-$l_n$ 
representations of $\cY (sl_2)$ 
$\cH_{1\dots N}=V_{2l_1}(\de_1)\otimes\dots\otimes V_{2l_N}(\de_N)$,
let us consider the general periodic inhomogeneous XXX Heisenberg
chain of length $N$ whose quantum space of states is $\cH_{1\dots N}$.

The quantum $L$-operator $L_n(u)$ at site $n$, which is linear in the spectral 
parameter $u$, is constructed in terms of the spin operators $\S_n^+$, $\S_n^-$
and $\S_n^z$ belonging to the irreducible finite dimensional representation
$V_{2l_n}$ (dim$V_{2l_n}=2l_n+1$) of the Lie algebra $sl_2$, 
and depends on the inhomogeneity parameter $\de_n$:
\begin{equation*}
L_n(u-\de_n) = u-\de_n +\eta \sum_{\alpha=1}^3 \S_n^\alpha \sigma^\alpha
       = \begin{pmatrix}
            u-\de_n+\eta \S_n^z & \eta \S_n^-\\
            \eta \S_n^+         &  u-\de_n-\eta \S_n^z
          \end{pmatrix},      
\end{equation*}
with
\begin{xalignat*}{2}
 \S_n^\pm &= \S_n^x \pm i\S_n^y, &
  [\S_n^+,\S_n^-] &= 2 \S_n^z,\\ 
  \S_n^2 &= 
     \frac12 (\S_n^+ \S_n^- + \S_n^- \S_n^+) + (\S_n^z)^2 = l_n (l_n+1), &
  [\S_n^z,\S_n^\pm] &= \pm \S_n^\pm.
\end{xalignat*}
The monodromy matrix of the chain,
\begin{equation}
  T(u)\equiv T_{1\dots N}(u;\de_1,\dots,\de_n) 
    = L_N (u-\delta_N)\dots L_1(u-\delta_1)
        = \begin{pmatrix}
              A(u) & B(u)\\
              C(u) & D(u)
          \end{pmatrix},
\end{equation}
is therefore a polynomial  of degree $N$ in the spectral parameter $u$:
\begin{equation}
  T(u) = \sum_{i=0}^N u^i 
                           \begin{pmatrix}
                             \mathcal{A}_i & \mathcal{B}_i \\
                             \mathcal{C}_i & \mathcal{D}_i
                           \end{pmatrix},
\end{equation}
where $\mathcal{A}_i$, $\mathcal{B}_i$, $\mathcal{C}_i$ and $\mathcal{D}_i$ 
are quantum operators acting on the total
quantum space $\cH$ of the chain.
Commutation relations of operators $A(u)$, $B(u)$, $C(u)$ and $D(u)$ are
given by the relation:
\begin{equation}
 \mathsf{R} (u-v)\, (T(u)\otimes\Id)\,(\Id\otimes T(v)) 
         = (\Id\otimes T(v))\,(T(u)\otimes\Id)\, \mathsf{R} (u-v),
\end{equation}
where $\mathsf{R}$ is the rational $R$-matrix associated to the fundamental 
representation of $\cY (sl_2)$:
\begin{equation*}\label{mat-R}
   \mathsf{R} (u)=
      \begin{pmatrix}
         1 & 0 & 0 & 0\\
         0 & b(u) & c(u) & 0\\
         0 & c(u) & b(u) & 0\\
         0 & 0 & 0 & 1
      \end{pmatrix}
  \qquad\qquad
     \begin{aligned}
      b(u)  &= \frac{u}{u+\eta},\\
      c(u)  &= \frac{\eta}{u+\eta}.
     \end{aligned}
\end{equation*}
%
%

%

As for the generalized $R$-matrices $R^\sg$ defined on $\End \cH$ as in
proposition~\ref{prop:R} from the
different $R$-matrices $R_{i j} \in V_{2 l_i}\otimes V_{2 l_j}$, 
they satisfy, for any element $\sg \in \perm_n$,
\begin{equation}
     R_{1\dots N}^{\sigma}\, T_{1\dots N} = 
            T_{\sigma(1)\dots\sigma(N)}\, R_{1\dots N}^{\sigma}, 
          \label{eq:r0qsa}
\end{equation}
where $T_{\sigma(1)\dots\sigma(N)} \in \End \cH$ is merely obtained by 
permutation of the ordered product of $L$-operators:
\begin{equation}\label{Tsigma}
  T_{\sigma(1)\dots\sigma(N)}(u) 
    = L_{\sg(N)} (u-\delta_{\sg(N)})\dots L_{\sg(1)}(u-\delta_{\sg(1)}).
\end{equation}
This means that, if these $R$-matrices admit factorizing $F$-matrices,
the latter induce
a basis of $\cH$ in which the expression of the monodromy matrix is
symmetric under any permutation~\eqref{Tsigma} of the sites.
In~\cite{MaiS96}, such a basis has been exhibited for spin $1/2$, which 
happened to diagonalize $D$. In the following, $F$-basis for higher spin chains
will be investigated directly as diagonalizing basis for the operator $D$.

\section{Functional Bethe Ansatz for the chain}

In this section, we apply to operator $D_{1\dots N}(u)$ 
the Functional Bethe Ansatz method of Sklyanin 
in order to obtain explicit expressions for operators $A$, $B$, $C$, $D$
in a basis in which $D$ is diagonal. We merely give here the idea of this
method applied to our special case, for the procedure is very similar
to what is done in~\cite{Skl90,Skl92}.

\bigskip

The idea of Functional Bethe Ansatz, applied here to operator $D$ 
instead of $B$, is to define the operator roots $\hx_n$, $1\le n \le N$ of the
polynomial of degree $N$ $D_{1\dots N}(u)$:
\begin{equation}\label{Dxn}
   D_{1\dots N}(u)= \prod_{n=1}^N (u-\hx_n),
\end{equation}
by constructing an isomorphism between $\Fun \Y$ and the space of states $\cH$
of the chain. Here $\Y \subset \C^N$ is the common spectrum of
$\{\hx_1,\dots,\hx_N\}$, and $\hx_n$ are realized on
$\Fun \Y$ as the operators of multiplication by the corresponding coordinates
(the $n$-th coordinate) in $\C^N$ (see~\cite{Skl90} for details):
\begin{equation*}
   \big[ \hx_n f \big] (\mathbf{y})= y_n \, f(\mathbf{y}),
    \quad \forall f \in \Fun\Y,\quad \forall \mathbf{y}=(y_1,\dots,y_N) \in \Y.
\end{equation*}

In our case, the spectrum $\Y$ can be determined directly from the
diagonal elements
of the lower triangular matrix $D_{1\dots N}(u)$.
Indeed, one can prove similarly as in~\cite{MaiS96} that the diagonal
of $D(u)$ is given by  
\[
\Diag D(u) = \prod\limits_{i=1}^N (u-\de_i-\eta \S^z_i),
\]
so that the associated spectrum is
\begin{equation}
   \Y=\La_1 \times \La_2 \times \dots \times \La_N,
    \quad \La_i=\{\de_i+\eta k \, |\, k=-l_i, -l_i+1, \dots, l_i\}.
\end{equation}

\begin{rem}
  FBA can be applied to $D(u)$ provided some conditions are satisfied (see
  \cite{Skl90}). They can be reduced here to the following 
  non-intersection condition on the 
  spectrum $\Y$:
  \begin{equation}
     \La_i \cap \La_j = \emptyset\quad \text{for}\ i\ne j.
  \end{equation}
  This ensures in particular that $D(u)$ is diagonalizable (since its spectrum
  is simple).
\end{rem}

Operators $\hX_n^+$, $\hX_n^-$ are then defined from polynomials $B(u)$
and $C(u)$ by substitution ``from the left'' (i.e. with some operator
ordering)  of $u$ by $\hx_n$:
\begin{equation}
  \hX_n^- = \sum_{p=0}^N \hx_n^p \mathcal{B}_p 
                     \equiv \left[ B(u)\right]_{u=\hx_n},\qquad
  \hX_n^+ = \sum_{p=0}^N \hx_n^p \mathcal{C}_p 
                     \equiv \left[ C(u)\right]_{u=\hx_n}.
\end{equation} 
The commutation relations for $\hx_n,\ \hX_m^\pm$ follow from those of
$A$, $B$, $C$, and $D$:
\begin{xalignat*}{2}
    &[\hx_m,\hx_n]=[\hX_m^\pm,\hX_n^\pm]=0,\quad\forall m,n, &
    &[\hX_m^+,\hX_n^-]=0,\quad\forall m,n,\ m\ne n,\\
    &\hX_m^\pm \hx_n=(\hx_n\mp\eta\de_{mn})\hX_m^\pm,\quad\forall m,n, & 
    &\hX_n^\mp \hX_n^\pm = - \Delta (\hx_n\pm\frac\eta 2),\quad\forall n,
\end{xalignat*}
where $\Delta(u)=\prod_{n=1}^N
(u-\de_n-l_n\eta-\frac \eta 2)(u-\de_n+l_n\eta+\frac \eta 2)$
is the quantum determinant of the monodromy matrix $T(u)$.

Conversely, $B(u)$ and $C(u)$ can be reconstructed in terms
of the $\hx_n,\ \hX_m^\pm$ by means of polynomial interpolation:
\begin{equation}
    B(u)  = \sum_{n=1}^N \bigg\{ \prod_{i \ne n} \frac{u-\hx_i}{\hx_n-\hx_i}
                 \bigg\}\,     \hX_n^-,\qquad
    C(u) = \sum_{n=1}^N \bigg\{ \prod_{i \ne n} \frac{u-\hx_i}{\hx_n-\hx_i}
                 \bigg\}\,     \hX_n^+,
\end{equation}
which, with~\eqref{Dxn} and the fact that $A$ can be obtained in terms of
$B$, $C$, $D$ and the quantum determinant $\Delta$, provides expressions
of the matrix elements of the monodromy matrix only in terms of operators
$\hx_n$ and $\hX_n^\pm$.
Therefore, if we compute explicitly the action of $\hX_n^\pm$ in a basis which
diagonalizes simultaneously all $\hx_n$, we will obtain explicit expressions
for $A$, $B$, $C$, and $D$ in a basis in which $D$ is diagonal.

This can be done by considering the 
actions of $\hx_n,\ \hX_m^\pm$ in an explicit
basis of $\Fun \Y$ (which is isomorphic to $\cH$). The point is thus to 
determine the action of $\hX_n^\pm$ on $\Fun \Y$. 

Still following Sklyanin,
let us define the function $\Delta_n^\pm = \hX_n^\pm \omega$, where 
$\omega$ is the constant function $\omega \equiv 1$ of $\Fun \Y$.
The action of $\hX_n^\pm$ on an arbitrary function $f \in \Fun \Y$ is then
given in terms of $\Delta_n^\pm$ (see~\cite{Skl90}):
\begin{equation}
  \big[ \hX_n^\pm f \big] (\mathbf{y}) 
   = \Delta_n^\pm (\mathbf{y})\, 
     f(E_n^{\mp} \mathbf{y}) 
  \quad \forall \mathbf{y} \in \Y,
\end{equation}
where $E_n^{\mp}$ are the shift operators in $\mathbb{C}^N$:
\[
   E_n^{\mp}: (y_1,\dots,y_n,\dots,y_N) \longrightarrow
                 (y_1,\dots,y_n\mp\eta,\dots,y_N).
\]  
It is thus sufficient for our purpose to compute $\Delta_n^\pm$.

The same way as in \cite{Skl90}, 
commutation relations for $\hx_n,\ \hX_m^\pm$ 
impose some conditions on $\Delta_n^\pm$, and
it is easy to see that these conditions determine $\Delta_n^\pm$ up to 
multiplication by an arbitrary function:
\begin{equation}
   \Delta_n^\pm (\mathbf{y})=\xi_\pm 
       \frac{\rho(\mathbf{y})}{\rho(E_n^{\mp}\mathbf{y})} \Delta^\pm (y_n)
   \quad \text{with}\quad 
   \Delta^\pm (y_n)= \prod_{i=1}^N (y_n-\de_i\pm\eta l_i),
\end{equation}
where $\rho$ is an arbitrary function having no zeroes on $\Y$, and
$\xi_+$, $\xi_- \in \mathbb{C}$ are such that $\xi_+ \xi_- =1$.
%
%
Note that the indetermination on $\rho$ corresponds to the 
indetermination on the isomorphism between $\cH$ and $\Fun \Y$.
In the following we choose $\rho$ to be the constant function 
$\rho(\mathbf{y})\equiv 1$,
which corresponds to fixing this isomorphism. 
Hence, in the basis
\begin{equation}
   \{f_{1\,k_1}\times f_{2\,k_2}\times\dots\times f_{N\,k_N} \,
     |\, f_{i\,k_i}\in\Fun\La_i,\  k_i=l_i, l_i-1, \dots, -l_i \}
\end{equation}
of $\Fun\Y$ defined by
\begin{equation}
   f_{n\,k_n}(\de_n+\eta p_n)=\xi_+^{-l_n+k_n}\, \de_{k_n,p_n}
   \quad \forall p_n\in \{ -l_n,\dots,l_n\},
\end{equation}
$\hx_n,\ \hX_n^\pm$ are given by
\begin{equation*}
   \hx_n = \de_n+\eta \sz_n,\qquad 
   \hX_n^\pm = \eta 
     \prod_{i=1 \atop i\ne n}^N (\de_n-\de_i+\eta\sz_n\pm\eta l_i)\, S_n^\pm,
\end{equation*}
where $\sz_n$, $S_n^\pm$ are the following matrix representations 
(dimension $2 l_n +1$) of
the spin operators $\S_n^z$, $\S_n^\pm$:
\begin{gather}
  S_n^z= 
  \begin{pmatrix}
    l_n    & 0      & \hdotsfor{2}    & 0\\
    0      & l_n-1  & 0      &\hdotsfor{1}  & 0\\
    \vdots & \ddots & \ddots & \ddots & \vdots\\
    0      & \hdotsfor{1}  &  0     & -l_n+1 & 0\\
    0      & \hdotsfor{2}    & 0      & -l_n
  \end{pmatrix}_{[n]},\\
  S_n^+= 
  \begin{pmatrix}
    0      & 2l_n     & 0 &\hdotsfor{1}  & 0\\
    \vdots & \ddots &\ddots      & \ddots &\vdots\\
    0      & \hdotsfor{1} & 0 & 2  & 0 \\
    0      & \hdotsfor{2}          & 0 & 1\\
    0      & \hdotsfor{3} &0
  \end{pmatrix}_{[n]},\quad  
  S_n^-= 
  \begin{pmatrix}
    0 & \hdotsfor{3}  & 0\\
    1 & 0 & \hdotsfor{2}  & 0\\
    0 & 2 & 0 &\hdotsfor{1} & 0\\
    \vdots & \ddots &\ddots      & \ddots &\vdots\\
    0 & \hdotsfor{1} & 0 & 2 l_n & 0
  \end{pmatrix}_{[n]}.
\end{gather}
The subscript $[n]$ means here that the corresponding operator acts as identity
on all spaces of the tensor product but $\Fun\La_n\simeq V_{2l_n}$.

As a consequence, we have the following proposition:
\begin{prop}
  There exists a basis of $\cH$ in which
  $D$, $B$, $C$ have the following expressions $\Dt$, $\Bt$ and $\Ct$:
  \begin{align}
    \Dt_{1\dots N}(u) &= \prod_{n=1}^N (u-\de_n-\eta S_n^z),\label{Dt}\\
    \Bt_{1\dots N}(u) &= \sum_{n=1}^N \bigg\{ \prod_{i\ne n} 
            (u-\de_i-\eta S_i^z)
            \frac{\de_n-\de_i+\eta \sz_n-\eta l_i}
                 {\de_n-\de_i+\eta \sz_n-\eta \sz_i} \bigg\}\,
            \eta \sm_n,\label{Bt}\\
    \Ct_{1\dots N}(u) &= \sum_{n=1}^N \bigg\{ \prod_{i\ne n} 
            (u-\de_i-\eta S_i^z)
            \frac{\de_n-\de_i+\eta \sz_n+\eta l_i}
                 {\de_n-\de_i+\eta \sz_n-\eta \sz_i} \bigg\}\,
            \eta \sp_n,\label{Ct}
  \end{align}
  $\At$ being given by the quantum determinant:
  \begin{equation}\label{At}
     \At(u) = \Dt^{-1}(u-\eta) 
        \big[ \Delta(u-\frac \eta 2)+\Bt(u-\eta) \Ct(u) \big].  
  \end{equation}
\end{prop}

These expressions are to be compared 
to the expressions for operators $B$, $C$, $D$, $A$ in the $F$-basis
obtained by J.M. Maillet and J. Sanchez de Santos in \cite{MaiS96}.
Indeed, in the case when all spins $l_i=1/2$, these formulas coincide
(up to a normalization factor).
Moreover, as in \cite{MaiS96},
$\Dt$ is still here a pure tensor product of diagonal matrices
and  the expressions for $\Dt$, $\Bt$, $\Ct$, and thus $\At$, are completely
symmetric under any permutation of the sites.
So, the new basis obtained by Functional Bethe Ansatz technique
appears to be a generalization of the $F$-basis obtained in \cite{MaiS96}.
In order to prove it, we will thus compute the matrix which induces
this change of basis, and show that it is effectively a factorizing $F$-matrix
in the sense of definition~\ref{def:factfm}.

\section{Determination of the $F$-matrix}
\label{sec:calculF}

In this section, we compute the matrix $F_{1\ldots N}$ 
which induces the previous change
of basis for the chain of length $N$, that is such that, for $X=A$, $B$, $C$, 
or $D$,
\begin{equation}
  \Xt_{1\ldots N}(u) = F_{1\ldots N} X_{1\ldots N}(u) F_{1\ldots N}^{-1},
\end{equation}
with $\At$, $\Bt$, $\Ct$ and $\Dt$ given by \eqref{Dt}-\eqref{At},
and we show that the matrices $F_{1\ldots N}$ thus defined 
(for any spin chains) are 
factorizing $F$-matrices.

\begin{rem}
   $F_{1\ldots N}$, which diagonalizes the lower triangular matrix 
   $D_{1\ldots N}$ whose diagonal coefficients are all distinct, 
   is itself lower triangular.
\end{rem}

\bigskip

In order to compute $F_{1\ldots N}$ by induction on $N$, let us define,
for all integer $n\ge 2$, 
\begin{equation}\label{mat:Ft-part}
  \Ft_{1,2\ldots n}=F_{1\ldots n}\, F_{2\ldots n}^{-1}.
\end{equation}
By definition, $\Ft_{1,2\ldots n}$ is a lower triangular matrix, and thus has
to be of the form
\begin{equation}
  \Ft_{1,2\ldots n}=\cQt_{1,2\ldots n} 
     \biggl(1 + \sum_{k=1}^{2l_1} \alpha_{1,2\ldots n}^{(k)}\,
               (S_1^-)^k \biggr),
\end{equation}
where $\cQt_{1,2\ldots n}$ and $\alpha_{1,2\ldots n}^{(k)}$ for all $k$ are
diagonal on space (1).

Using the relations between $A$, $B$, $C$ and $D$ for $N$ and $N-1$ sites,
\begin{align}
  D_{1\ldots N}(u) &=\eta S_1^-\, C_{2\ldots N}(u) 
                + (u-\delta_1 - \eta S_1^z)\, D_{2\ldots N}(u),\\
  B_{1\ldots N}(u) &=\eta S_1^-\, A_{2\ldots N}(u) 
                + (u-\delta_1 - \eta S_1^z)\, B_{2\ldots N}(u),\\   
  C_{1\ldots N}(u) &=(u-\delta_1+\eta S_1^z)\, C_{2\ldots N}(u) 
                + \eta S_1^+\, D_{2\ldots N}(u),
\end{align}
and applying
the change of basis induced by $F_{1\ldots N}=\Ft_{1,2\ldots N} F_{2\ldots N}$,
one obtains the following equations for $\Ft_{1,2\ldots N}$:
\begin{align}
  \Dt_{1\ldots N}(u)\, \Ft_{1,2\ldots N} &=\Ft_{1,2\ldots N} 
                \left[ \eta S_1^-\, \Ct_{2\ldots N}(u) 
                + (u-\delta_1 - \eta S_1^z)\, \Dt_{2\ldots N}(u) \right],
  \label{eq:D}\\
  \Bt_{1\ldots N}(u)\, \Ft_{1,2\ldots N} &=\Ft_{1,2\ldots N} 
                \left[ \eta S_1^-\, \At_{2\ldots N}(u) 
                + (u-\delta_1 - \eta S_1^z)\, \Bt_{2\ldots N}(u) \right],
  \label{eq:B}\\   
  \Ct_{1\ldots N}(u)\, \Ft_{1,2\ldots N} &=\Ft_{1,2\ldots N}
                \left[ (u-\delta_1+\eta S_1^z)\, \Ct_{2\ldots N}(u) 
                + \eta S_1^+\, \Dt_{2\ldots N}(u) \right].
  \label{eq:C}
\end{align}

These {\em linear} equations enable us to compute $\Ft_{1,2\ldots N}$.
Indeed, decomposing~\eqref{eq:D} on all $(S_1^-)^k$, $0\le k\le 2l_1$,
we obtain:
\begin{equation}\label{Ft-par}
  \Ft_{1,2\ldots N}=\cQt_{1,2\ldots N} 
      \sum_{k=0}^{2l_1} \frac{1}{k!} 
             \big[\Ct_{2\ldots N}(\delta_1+\eta S_1^z)\,
              \Dt_{2\ldots N}^{-1}(\delta_1+\eta S_1^z)\big]^k  (S_1^-)^k ,
\end{equation}
where  $\cQt_{1,2\ldots N}$, which has to commute with
$\Dt_{1\ldots N}(u)$ for all values of $u$, is necessarily a diagonal 
matrix.
It is determined, up to a global numerical factor, using~\eqref{eq:B} 
and \eqref{eq:C}:
\begin{align}
  \cQt_{1,2\ldots N} &=\alpha (S_1^z; S_2^z,\ldots,S_N^z),\nonumber\\
                    &=\prod_{k=1}^{\infty} \left\{ 
     \frac{\Dt_{2\dots N}(\de_1+\eta\sz_1+\eta k)}
          {\Dt_{2\dots N}(\de_1+\eta l_1+\eta k)}\,
     \prod_{i=2}^N \frac{\de_1-\de_i+\eta l_1+\eta l_i+\eta k}
                        {\de_1-\de_i+\eta\sz_1+\eta l_i+\eta k} \right\} \,
      \al(l_1; l_2,\ldots,l_N).
\label{cD1}
\end{align}
In the following, we choose the normalization $\al(l_1; l_2,\ldots,l_N)=1$ 
so that $\Ft_{1,2\ldots N}\ket{0}=\ket{0}$ where $\ket{0}=(1,0,\dots,0)$.
Note that the product is actually finite for each matrix element of $\sz_1$.

The matrix $F_{1\ldots N}$ inducing the change of basis is thus given by 
induction on the number of sites $N$ of the chain:
\begin{align}\label{Fn}
  & F_N = 1,\\
  & F_{12\ldots N}=\Ft_{1,2\ldots N}\, F_{2\ldots N},\label{recurrence}
\end{align}
with $\Ft_{1,2\ldots N}$ given by \eqref{Ft-par}-\eqref{cD1}.

Its inverse can be computed by means of the linear equation
for $\Ft_{1,2\dots N}^{-1}$ equivalent to~\eqref{eq:D}. One obtains:
\begin{equation}
  \Ft_{1,2\ldots N}^{-1}=\bigg\{ \sum_{k=0}^{2l_1} \frac{(-1)^k}{k!} (S_1^-)^k
              \big[\Dt_{2\ldots N}^{-1}(\delta_1+\eta S_1^z)\,
              \Ct_{2\ldots N}(\delta_1+\eta S_1^z)\big]^k \bigg\}\, 
                 \cQt_{1,2\ldots N}^{-1}.  
\end{equation}

\bigskip

Note that the set of matrices $F_{1\dots N}$ can also be determined similarly 
by computing the other  
partial matrices
\begin{equation}\label{recurrence'}
 \Ft_{1\ldots n-1,n} = F_{1\ldots n}\, F_{1\ldots n-1}^{-1}.
\end{equation}
Linear equations similar to \eqref{eq:D}-\eqref{eq:C} lead to the
following expressions for these partial matrices:
\begin{align}\label{Ft'}
  \Ft_{1\ldots N-1, N} &=\cQt_{1\ldots N-1,N} 
      \sum_{k=0}^{2l_N} \frac{(-1)^k}{k!} 
          \big[\Bt_{1\ldots N-1}(\delta_N+\eta S_N^z) \,
           \Dt_{1\ldots N-1}^{-1}(\delta_N+\eta S_N^z)\big]^k  (S_N^+)^k ,\\
  \Ft_{1\ldots N-1, N}^{-1} &=\bigg\{ \sum_{k=0}^{2l_N} \frac{1}{k!} (S_N^+)^k
          \big[\Dt_{1\ldots N-1}^{-1}(\delta_N+\eta S_N^z)\, 
            \Bt_{1\ldots N-1}(\delta_N+\eta S_N^z)\big]^k \bigg\}\, 
                     \cQt_{1\ldots N-1,N}^{-1} ,
\end{align}
with
\begin{equation}
  \cQt_{1\ldots N-1,N} =\prod_{k=1}^{\infty}\left\{ 
     \frac{\Dt_{1\dots N-1}(\de_N+\eta\sz_N-\eta k)}
          {\Dt_{1\dots N-1}(\de_N-\eta l_N-\eta k)}\,
      \prod_{i=1}^{N-1} \frac{\de_N-\de_i-\eta l_N-\eta l_i-\eta k}
                             {\de_N-\de_i+\eta \sz_N-\eta l_i-\eta k} \right\}.
\label{cD2}
\end{equation}

It has also to be mentioned that other sets of partial $F$-matrices, defined
as $F_{1,2\dots n}=F_{2\dots n}^{-1} F_{1\dots n}$ and
$F_{1\dots n-1, n}=F_{1\dots n-1}^{-1} F_{1\dots n}$
are obtained from  \eqref{Ft-par}-\eqref{cD1} and \eqref{Ft'}-\eqref{cD2}
by replacing in these formulas 
$\Ct_{2\dots N}$, $\Dt_{2\dots N}$ and $\Bt_{1\dots N-1}$, $\Dt_{1\dots N-1}$
respectively by $C_{2\dots N}$, $D_{2\dots N}$ and $B_{1\dots N-1}$, 
$D_{1\dots N-1}$.

\begin{rem}
   In particular, $F_{12}$ is given by
  \begin{equation}\label{F12}
    F_{12}= \cQ_{12} \sum_{k=0}^{\infty} \frac{\eta^k}{k!}
                \prod_{j=1}^k [\de_1-\de_2+\eta\sz_1-\eta\sz_2+\eta j]^{-1}
                               (\sm_1)^k (\sp_2)^k, 
  \end{equation}  
   and its inverse is
  \begin{equation}\label{F12inv}
    F_{12}^{-1}= \bigg\{ \sum_{k=0}^{\infty} \frac{(-\eta)^k}{k!}
                 (\sm_1)^k (\sp_2)^k
                 \prod_{j=1}^k [\de_1-\de_2+\eta\sz_1-\eta\sz_2-\eta j]^{-1}
                               \bigg\} \, \cQ_{12}^{-1}.
  \end{equation}
\end{rem}

\begin{rem}
  The non-diagonal part of the matrix $F_{12}^{-1}$, which satisfies the linear
  equation
  \begin{equation}
     F_{12}^{-1}\, \Dt_{12}=D_{12}\,F_{12}^{-1},
  \end{equation}
  can be directly obtained from it as an infinite formal product:
  \begin{equation}
     F_{12}^{-1}=\prod_{k=0}^{\longrightarrow \atop \infty} 
      \Dt_{12}^k(u) \, D_{12}(u) \, \Dt_{12}^{-(k+1)}(u)\ \cdot\ \cQt_{12}^{-1}.
  \end{equation}
  Note that by computing this product explicitly, one finds again the 
  expression~\eqref{F12inv} (in particular $F_{12}^{-1}$ does not depend
  on the spectral parameter $u$).
\end{rem}

\bigskip

\begin{thm}\label{thmF}
  The matrices $F_{1\ldots N}$ given by induction on $N$ by 
  \eqref{Fn}, \eqref{recurrence},
  \eqref{Ft-par} and \eqref{cD1}
  provide a set of factorizing $F$-matrices in the sense of 
  definition~\ref{def:factfm}.
\end{thm}

\Proof
The matrices $F_{1\ldots N}$, which induce the change of basis~\eqref{Bt},
\eqref{Ct}, \eqref{Dt}, \eqref{At}, being invertible by definition,
we merely have to show that, for any permutation $\sigma \in \perm_N$,
\begin{equation*}
  F_{\sigma(1)\ldots\sigma(N)}\, R_{1\ldots N}^\sigma 
  = F_{1\ldots N}.
\end{equation*}  
For $X= A,\ B,\ C$, or $D$, one knows from~\eqref{eq:r0qsa} that
\begin{equation*}
  R_{1\ldots N}^\sigma \, X_{1\ldots N} 
  = X_{\sigma(1)\ldots\sigma(N)} \, R_{1\ldots N}^\sigma.
\end{equation*}
The formulas for 
$\Xt_{1\ldots N}= F_{1\ldots N} X_{1\ldots N} F_{1\ldots N}^{-1}$ are given
by \eqref{Dt}, \eqref{Bt}, \eqref{Ct} and \eqref{At}, and are completely 
symmetric under any permutation of the sites:
\begin{equation*}
  \Xt_{\sigma(1)\ldots\sigma(N)}=\Xt_{1\ldots N}, 
       \quad \forall \sigma \in \perm_N.
\end{equation*}
Thus
\begin{equation*}
  R_{1\ldots N}^\sigma \, 
 (F_{1\ldots N}^{-1} \, \Xt_{1\ldots N} \, F_{1\ldots N}) 
  = (F_{\sigma(1)\ldots\sigma(N)}^{-1} \, \Xt_{\sigma(1)\ldots\sigma(N)} \,
     F_{\sigma(1)\ldots\sigma(N)}) \, R_{1\ldots N}^\sigma, 
\end{equation*} 
which implies that the quantity
$F_{\sigma(1)\ldots\sigma(N)} \, R_{1\ldots N}^\sigma \, F_{1\ldots N}^{-1}$
commutes with $\At(u)$, $\Bt(u)$, $\Ct(u)$ and $\Dt(u)$ for all the values of
$u$. Therefore, this is equal to the identity matrix times 
a numerical factor, which is $1$ for the appropriate
normalization of $R$.
\qed

\begin{rem}\label{Rgauss}
  The expression for $F_{21}^{-1} F_{12}$ which follows from \eqref{F12} and 
  \eqref{F12inv} coincides with the corresponding finite dimensional 
  representation for the Gauss decomposition of the universal $R$-matrix 
  of the Yangian double $\mathcal{D}\cY(sl_2)$ obtained in~\cite{KhoT96} by
  Khoroshkin and Tolstoy:
  \begin{equation}
     R_{12}=F_{21}^{-1} F_{12}= R_+\, R_0\, R_- ,
  \end{equation}
  with
  \begin{equation}
     \begin{aligned}
     R_+ &= \sum_{k=0}^{\infty} (\sp_1)^k (\sm_2)^k 
       \Big[ k! \prod_{j=1}^k (\lambda+\sz_1-\sz_2+j)\Big]^{-1},\\
     R_- &= \sum_{k=0}^{\infty} 
       \Big[ k! \prod_{j=1}^k (\lambda+\sz_1-\sz_2+j)\Big]^{-1}
                (\sm_1)^k (\sp_2)^k,\\
     R_0 &= \prod_{k=0}^{\infty}
       \frac{(\lambda+\sz_1-\sz_2+k)(\lambda-l_1-l_2+k)
             (\lambda+\sz_1-\sz_2+k+1)(\lambda+l_1+l_2+k+1)}
            {(\lambda-l_1-\sz_2+k)(\lambda+\sz_1-l_2+k)
             (\lambda+l_1-\sz_2+k+1)(\lambda+\sz_1+l_2+k+1)},
    \end{aligned}
  \end{equation}
  where $\lambda=\frac{\de_1-\de_2}{\eta}$.
  So the computation of $F$ leads to nice factorized expressions, in any 
  finite dimensional representation, 
  of the $R$-matrices associated to any permutation $\sg \in \perm_n$.
  In particular, this gives a hint concerning an universal formula for $F$:
  the non-diagonal term of $F$ in our formula corresponds exactly
  to the representation $R_-$ of the non-diagonal part of the Gauss 
  decomposition of the universal $R$-matrix of $\mathcal{D}\cY(sl_2)$, 
  and hence admits a
  universal formula; the question is thus if there exists an appropriate
  factorization of the diagonal part of this Gauss decomposition at the 
  universal level. 
\end{rem}

\begin{rem}
   Theorem~\ref{thmF} can also be easily proved by means of 
   proposition~\ref{prop:fundtf} using remark~\ref{Rgauss} and the explicit
   expression obtained for $\Ft_{1,2\dots N}$ and $\Ft_{1\dots N-1,N}$.
\end{rem}

\section{Conclusion}

In this letter we have computed the factorizing $F$-matrices representing
Drinfel'd twists for all finite dimensional evaluation representations
of Yangian $\cY(sl_2)$.
The FBA technique enables us to do this by solving only {\em linear}
equations.
The next step would be to generalize this to
non-finite dimensional representations and to
obtain an eventual universal form for this twist. Let us note here that,
in our formula, only the diagonal part depends on the dimension of the 
representation, whereas the non-diagonal terms simply correspond to
the representations
of the non-diagonal parts of the Gauss decomposition of the 
universal $R$-matrix given in~\cite{KhoT96}, and hence admit universal
formulas. The point would be here
to find an appropriate factorization of the diagonal part of this Gauss 
decomposition at the universal level, in order to obtain, as in~\cite{FreM92}, 
the universal
$R$-matrix as a product $(\cF_{21}^-)^{-1} \cF_{12}^+$, with $\cF^+$, $\cF^-$
satisfying the cocycle relation and $\cF^+=\cF^-$ only for finite dimensional
representations (in general, $\cR$ is not unitary, but pseudo-unitary).

The results obtained in this letter open furthermore 
the possibility to compute 
form factors for the XXX Heisenberg chain of spins $l$, by using the new basis
given by this $F$-matrix, in the spirit of what has been
done for spin-$\frac12$ in~\cite{KitMT98}.
Let also us mention here that the method we used to compute the factorizing
$F$-matrices is most probably applicable to Yangians or 
quantum affine algebras associated to higher rank Lie algebras.

\vspace{0.5cm}

{\large\bf{Acknowledgements.}}
I would like to thank J.-M. Maillet for many useful discussions and remarks.

\bibliographystyle{h-elsevier} 

\bibliography{biblio}

\begin{thebibliography}{10}

\bibitem{Dri85}
V.G. Drinfeld,
\newblock Soviet Math. Dokl. 32 (1985) 254.

\bibitem{Dri87}
V.G. Drinfel'd,
\newblock Proceedings of the the International Congress of Mathematicians,
  Berkeley, USA, 1986, pp. 798--820, AMS, 1987.

\bibitem{Jim85}
M. Jimbo,
\newblock Lett. Math. Phys. 10 (1985) 63.

\bibitem{Jim86}
M. Jimbo,
\newblock Lett. Math. Phys. 11 (1986) 247.

\bibitem{Dri83a}
V.G. Drinfel'd,
\newblock Soviet Math. Dokl. 28 (1983) 667.

\bibitem{Dri90a}
V.G. Drinfel'd,
\newblock Leningrad Math. J. 1 (1990) 1419.

\bibitem{MaiS96}
J.M. Maillet and J. Sanchez~de Santos,
\newblock (1996), q-alg/9612012.

\bibitem{KitMT98}
N. Kitanine, J.M. Maillet and V. Terras,
\newblock (1998), math-ph/9807020.

\bibitem{Skl90}
E.K. Sklyanin,
\newblock Functional {B}ethe ansatz,
\newblock Integrable and superintegrable systems, edited by B. Kupershmidt, pp.
  8--33, World Scientific, 1990.

\bibitem{Skl92}
E.K. Sklyanin,
\newblock Quantum group and Quantum Integrable Systems, edited by M.L. Ge, pp.
  63--97, Nankai Lectures in Mathematical Physics, World Scientific, 1992.

\bibitem{ChaP90}
V. Chari and A. Pressley,
\newblock L'Enseignement Math. 36 (1990) 267.

\bibitem{KhoT96}
S. Khoroshkin and V. Tolstoy,
\newblock Lett. Math. Phys. 36 (1996) 373.

\bibitem{FreM92}
L. Freidel and J.M. Maillet,
\newblock Phys. Lett. B 296 (1992) 353.

\end{thebibliography}


\end{document}